\documentclass{PoS}

\usepackage{graphics}
\usepackage{amsmath}
\usepackage{xcolor}
\usepackage{braket}
\usepackage{cases}
\usepackage{slashed}
\usepackage[utf8]{inputenc}
\usepackage[T1]{fontenc}
\usepackage{ulem}
\usepackage{subcaption}

\title{Relativistic approach to nuclear spin-isospin excitations including quasiparticle-vibration coupling}

\ShortTitle{Relativistic approach to nuclear spin-isospin excitations including quasiparticle-vibration coupling}

\author{\speaker{Caroline Robin} \\%
        Department of Physics, Western Michigan University, Kalamazoo, MI 49008, USA\\
        E-mail: \email{caroline.robin@wmich.edu}}

\author{Elena Litvinova\\
         Department of Physics, Western Michigan University, Kalamazoo, MI 49008, USA, and\\
         National Superconducting Cyclotron Laboratory, Michigan State University, East Lansing, MI 48824, USA \\
        E-mail: \email{elena.litvinova@wmich.edu}}

\abstract{The spin-isospin response of stable and exotic nuclei is investigated in the framework of the proton-neutron relativistic quasiparticle time-blocking approximation (pn-RQTBA). Based on the Covariant Density Functional Theory, this method extends the proton-neutron Relativistic Quasiparticle Random-Phase Approximation (pn-RQRPA) by including the coupling between single quasiparticles and collective nuclear vibrations. In the charge-exchange channel, this coupling generates a time-dependent effective interaction between proton and neutron quasiparticles. The particle-hole component of this interaction adds to the static pion and rho-meson exchange, while the particle-particle component provides a microscopic and consistent proton-neutron pairing interaction.
We find that such dynamical effects induce fragmentation and spreading of the Gamow-Teller transition strength which are important for a better agreement with the experimental measurements and for an accurate description of $\beta$-decay rates.
The new developments include the coupling of single nucleons to isospin-flip vibrations in doubly-magic nuclei. We find that these phonons can have a non-negligible effect on $\beta$-decay half-lives and "quenching" of the strength.}

\FullConference{The 26th International Nuclear Physics Conference\\
                 11-16 September, 2016\\
                 Adelaide, Australia}

\begin{document}

\section{Introduction}
Nuclear transitions involving transfer of isospin have many applications in nuclear and particle physics, as well as in astrophysics as they determine weak-interaction rates governing r-process nucleosynthesis and stellar evolution. 
Such types of modes as Fermi, Gamow-Teller (GT) or spin-dipole (SD) transitions are very sensitive to the spin-isospin dependence of the nucleon-nucleon interaction and have been extensively studied in many theoretical methods (see \textit{e.g.} Refs \cite{Brown} to \cite{Robin}).
In the Relativistic Random-Phase Approximation (RRPA) based on the Relativistic Mean-Field (RMF), where the exchange interaction is not explicitly treated, isospin-transfer modes are fully determined by isovector rho- and pion-exchange. While the proton-neutron (R)RPA (pn-(R)RPA) usually describes the position of GT and SD giant resonances quite well, extensions are necessary to include spreading mechanism essential for the description of width and details of the transition strength.
In open-shell nuclei with superfluid pairing correlations, it has been shown that an isoscalar residual pairing interaction is important for an accurate description of the GT strength and weak-interaction rates \cite{Engel,Mustonen,Bai,Niksic}. In proton-neutron (Relativistic) Quasiparticle-RPA (pn-(R)QRPA) studies, the strength of this interaction is however usually fitted to $\beta$-decay half-lives which lowers the predictive power of the method.\\
In this work we study the GT response of a few mid-mass nuclei within the Relativistic Quasiparticle-Vibration Coupling (RQVC) framework presented in Ref. \cite{Robin}.
This approach, applied in the proton-neutron Quasiparticle Time-Blocking Approximation (pn-RQTBA), extends the pn-RQRPA by accounting for the interplay between quasiparticle and collective degrees of freedom, and naturally generates a dynamical pn pairing interaction, without extra empirical factor.
We compare our results to the pn-RQRPA and to the available experimental data. For the first time we include the coupling of nucleons to isospin-flip vibrations in doubly-magic nuclei. We investigate their effect on the quenching of the GT strength and $\beta$-decay half-lives.

\section{Spin-isospin response of nuclei}
\begin{figure*}
\centering
\resizebox{0.8\textwidth}{!}{%
\includegraphics{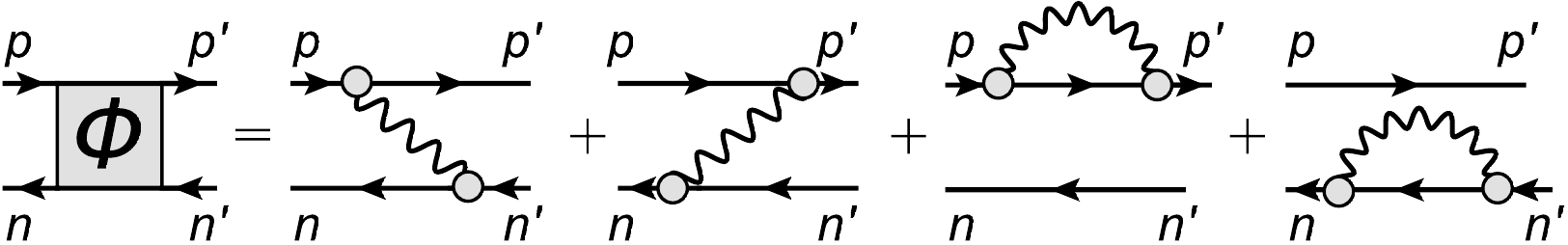}
}
\caption{Quasiparticle-Vibration Coupling interaction in the proton-neutron channel.}
\label{fig:Phi_IS}
\end{figure*}

The study of the response of nuclei to a weak external field $\hat F$ is a great tool to access their excitation spectra.
At first order in the perturbation, such field excites the nucleus from its ground state by inducing particle-hole excitations. 
The corresponding transition strength distribution $S(E)$ can be directly obtained from the knowledge of the propagator of a correlated particle-hole pair in the nuclear medium, or response function $R$, since
\begin{eqnarray}
S(E) = \sum_f |\braket{\Psi_f|\hat F|\Psi_i}|^2 \delta (E-E_f+E_i) 
        = -\frac{1}{\pi} \lim_{\Delta\rightarrow 0^+} \mbox{Im} \braket{\Psi_i| \hat{F}^\dagger R(E+i\Delta) \hat F |\Psi_i} \; ,
\label{eq:strength}
\end{eqnarray}
where $\ket{\Psi_i}$ and $\ket{\Psi_f}$ denote the nuclear ground and excited states respectively. In the framework of the linear response theory, the response function is the solution of the Bethe-Salpeter equation:
\begin{eqnarray}
\boldsymbol{R}(14,23) &=& \boldsymbol{G}(1,3) \boldsymbol{G}(4,2) -i \int d5... d8 \boldsymbol{G}(1,5) \boldsymbol{G}(6,2) \boldsymbol{V}(58,67) \boldsymbol{R}(74,83) \; ,\label{eq:BSE}
\end{eqnarray}
where the indices $i=1,2,3...\equiv(t_i,k_i)$ denote the time variable $t_i$ along with a complete set of nucleonic quantum numbers $k_i$. 
In open-shell nuclei with superfluid correlations, the single-nucleon space acquires an extra dimension and these indices are supplemented by an additional quantum number $\eta_i=\pm$ denoting the upper and lower component in the Nambu-Gorkov space.
In Eq. (\ref{eq:BSE}), $G$ denotes in principle the exact one-nucleon propagator, while $V$ is an effective two-body interaction induced by the medium. This interaction is consistently determined from the single-nucleon self-energy $\Sigma$ as
\begin{equation}
V(58,67) = i \frac{\delta \Sigma(5,6)}{\delta G(7,8)} \; .
\end{equation}
In the present relativistic quasiparticle-vibration coupling (RQVC) model the static part of the self-energy is determined within the relativistic mean-field approximation with pairing correlations, or relativistic Hartree-Bogoliubov (RHB) approximation, assuming a static meson-exchange interaction. This leads to a description of the ground state where the (quasi)nucleons evolve independently in classical meson fields. 
In reality, however, many virtual interaction processes can occur during the propagation of a nucleon in the nucleus. In particular, due to correlations, the excitation of a (quasi)particle-(quasi)hole (qp-qh) pair can induce the coherent excitation of all nucleons, leading to a vibration of the nucleus (phonon), which in turn, produces polarization of the medium and modification of the single-particle motion.
Such effects are taken into account by introducing an energy-dependent term to the self-energy, describing the emission and re-absorption of a nuclear vibration by a (quasi)particle.
Subsequently, the effective interaction in the BSE (\ref{eq:BSE}) is supplemented by the dynamical exchange of a phonon between two (quasi)particles.
In the charge-exchange channel, and in the time-blocking approximation (TBA) described in \cite{TBA,Litvinova2008}, the BSE for the qp-qh propagator takes the following form in the energy representation:
\begin{eqnarray}
R^{\eta_1 \eta_4 , \eta_2 \eta_3}_{k_{1_p} k_{4_n}, k_{2_n} k_{3_p}} (\omega) = \widetilde{R}^{(0) \eta_1 \eta_4 , \eta_2 \eta_3}_{k_{1_p} k_{4_n}, k_{2_n} k_{3_p}} (\omega)
+ \sum_{\substack{k_{5_p} k_{6_n} k_{7_p} k_{8_n} \\ \eta_5 \eta_6 \eta_7 \eta_8}} \widetilde{R}^{(0) \eta_1 \eta_6 , \eta_2 \eta_5}_{k_{1_p} k_{6_n}, k_{2_n} k_{5_p}} (\omega) 
W^{\eta_5 \eta_8 , \eta_6 \eta_7}_{k_{5_p} k_{8_n}, k_{6_n} k_{7_p}} (\omega) R^{\eta_7 \eta_4 , \eta_8 \eta_3}_{k_{7p} k_{4_n}, k_{8_n} k_{3_p}} (\omega) \; , \nonumber \\ \label{eq:BSE_pn}
\end{eqnarray}
where ($p,n$) refer to proton and neutron indices respectively, $\omega = E+i\Delta$ is the energy variable, $\widetilde{R}^{(0)}$ denotes the free qp-qh propagator in the proton-neutron channel, and $W$ is the effective interaction given by the sum of the static meson exchange and the energy-dependent amplitude $\Phi(\omega)$ containing the effect of QVC:
\begin{eqnarray}
W(\omega) = \widetilde V_\rho  + \widetilde V_\pi + \widetilde V_{\delta_{\pi}} +  \Phi(\omega)\; .
\label{eq:int}
\end{eqnarray}
\noindent The sum $\widetilde V_\rho  + \widetilde V_\pi$ is the finite range isovector meson-exchange interaction, while $\widetilde V_{\delta_{\pi}}$ denotes the zero-range Landau-Migdal term taken with parameter $g'=0.6$, as the exchange interaction is not treated in the present work \cite{Liang}. When considering this static interaction alone, one gets back to the pn-RQRPA.
The QVC amplitude $\Phi(\omega)$ is represented in Fig. \ref{fig:Phi_IS} in terms of Feynman diagrams. This dynamical interaction introduces 1(q)p-1(q)h$\otimes$phonon configurations and is responsible for damping of the transition strength.
A detailed expression of $W(\omega)$ can be found in Ref. \cite{Robin}.
In the following the extension of the pn-RQRPA including QVC effects in the TBA is referred to as pn-RQTBA.

\section{Gamow-Teller response of stable and exotic nuclei}
\begin{figure} [h!]
\centering
\subcaptionbox{GT$_-$ strength in $^{68}$Ni calculated with a smearing $\Delta=200$ keV. We also show the low-lying strength calculated with $\Delta=20$ keV. The arrow denotes the position of the $1^+_1$ state in $^{68}$Cu \cite{nndc}. \label{strength_68Ni_1}}%
{\includegraphics[width=.48\textwidth] {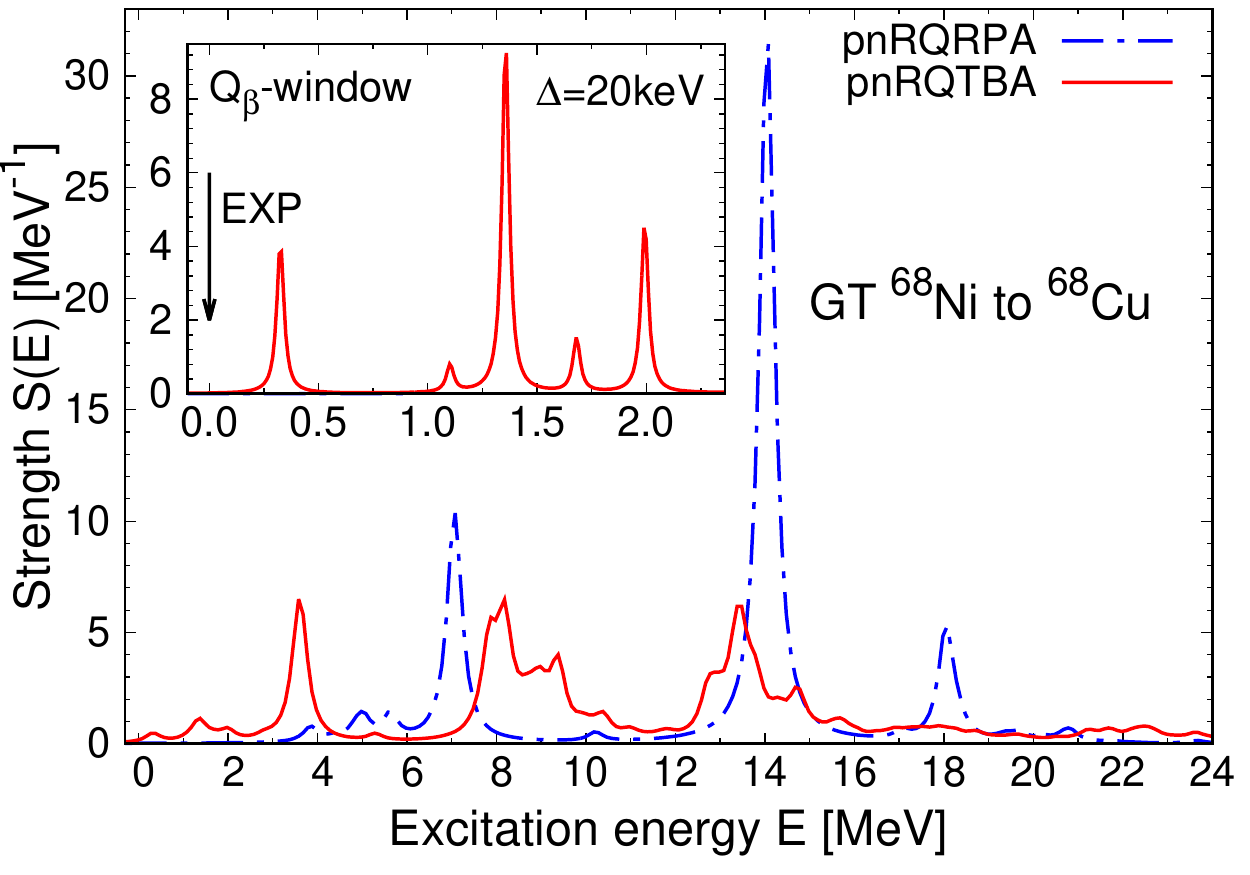}}\hfill
\subcaptionbox{GT$_-$ strength in $^{78}$Ni calculated with a smearing $\Delta=200$ keV. \label{strength_78Ni_1}}%
{\includegraphics[width=.48\textwidth] {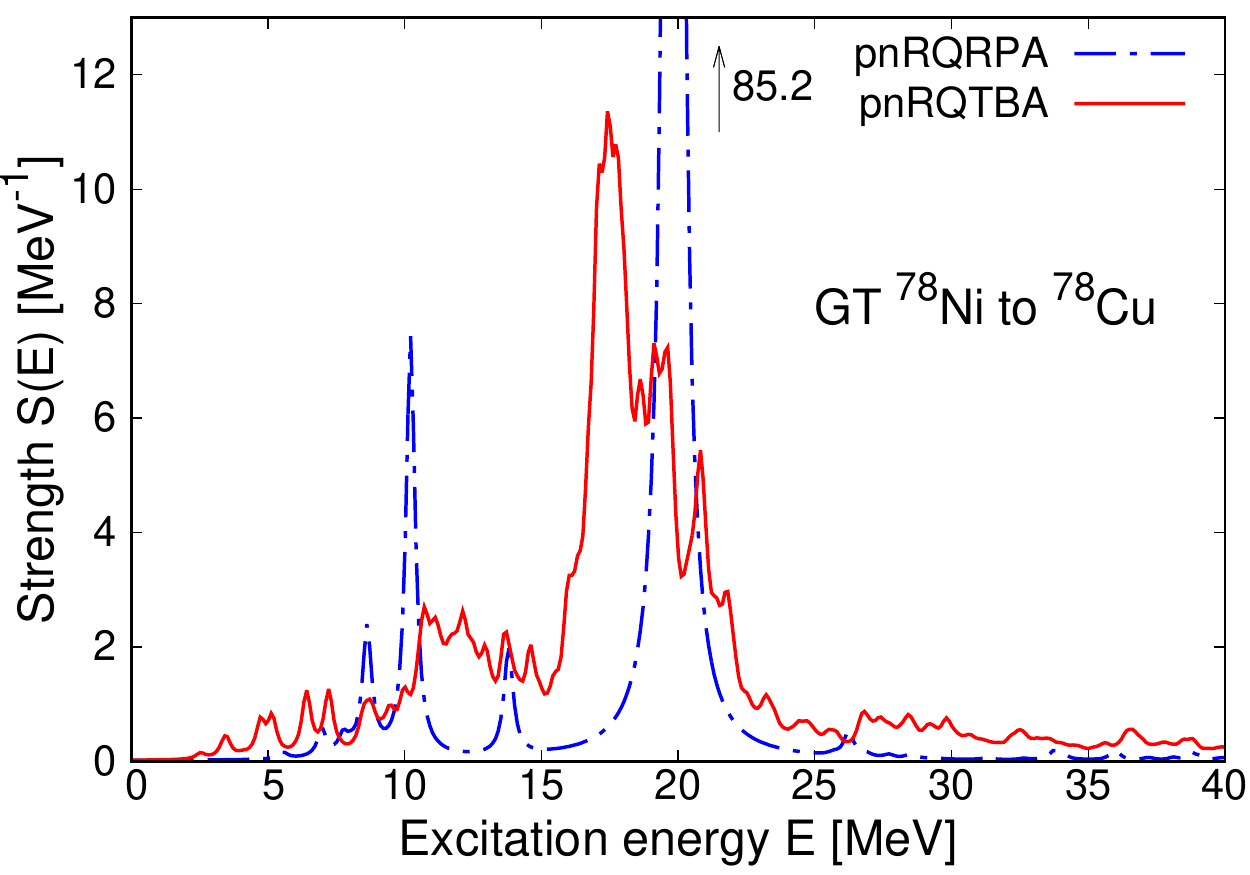}}  \hfill
\subcaptionbox{GT$_-$ strength in $^{90}$Zr calculated with a smearing $\Delta=1$ MeV. The experimental distribution has been extracted from Ref. \cite{Wakasa}. \label{strength_90Zr_1}}%
{\includegraphics[width=.48\textwidth] {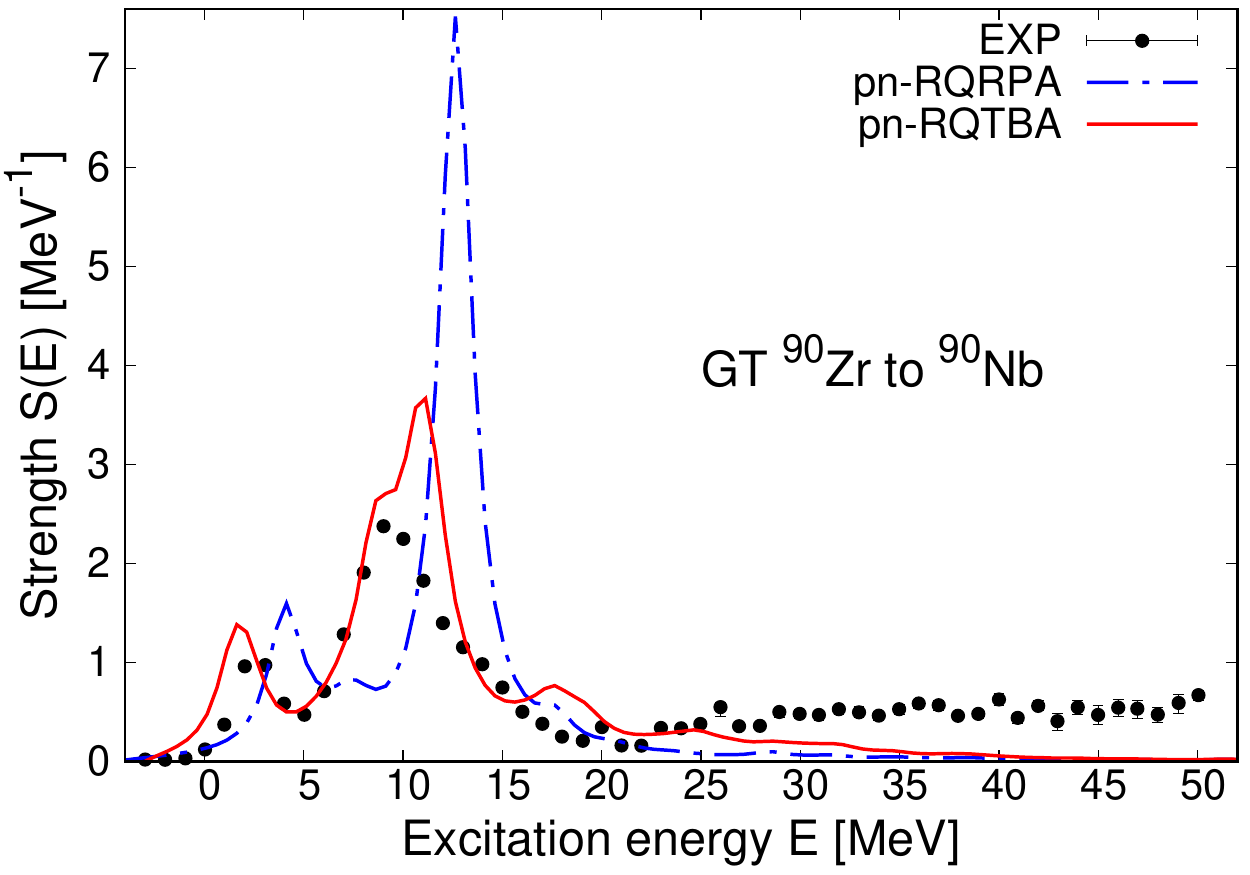}}  \hfill
\subcaptionbox{GT$_-$ strength in $^{48}$Ca calculated with a smearing $\Delta=210$ keV. The experimental distribution has been extracted from Ref. \cite{Yako}.  \label{strength_48Ca_1}}%
{\includegraphics[width=.48\textwidth] {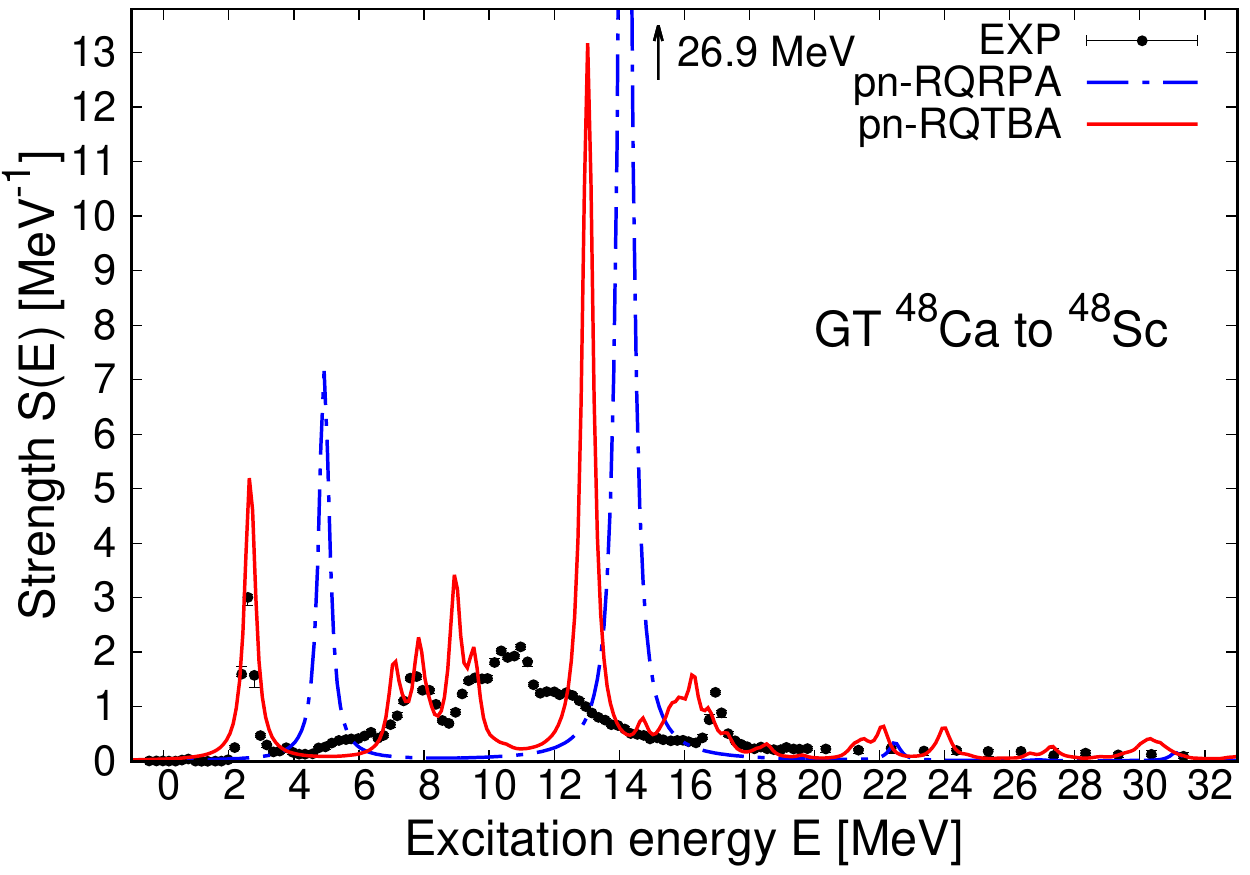}} \hfill
\caption{GT$_-$ strength distribution in a few stable and neutron-rich nuclei. The pn-RQRPA results are shown with blue lines while the pn-RQTBA results (with QVC) are shown with red lines.}
\label{fig:GT_1} 
\end{figure}

\begin{figure*}
\centering
\resizebox{\textwidth}{!}{%
\includegraphics{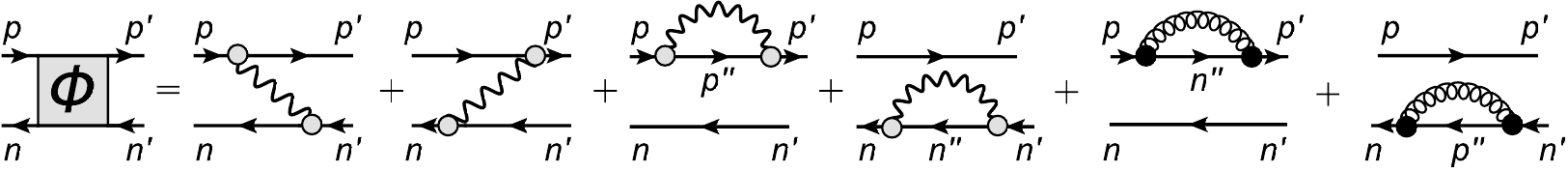}
}
\caption{QVC interaction in the proton-neutron channel. The wavy lines represent isoscalar (non-isospin-flip) phonons while the springs represent isovector (isospin-flip) phonons.}
\label{fig:Phi_IS+IV}
\end{figure*}

\begin{figure}[h!]
\centering
\subcaptionbox{$^{78}$Ni GT$_-$ transition strength to the Dirac sea of antiparticles. \label{strength_78Ni_Dirac}}%
{\includegraphics[width=.49\textwidth] {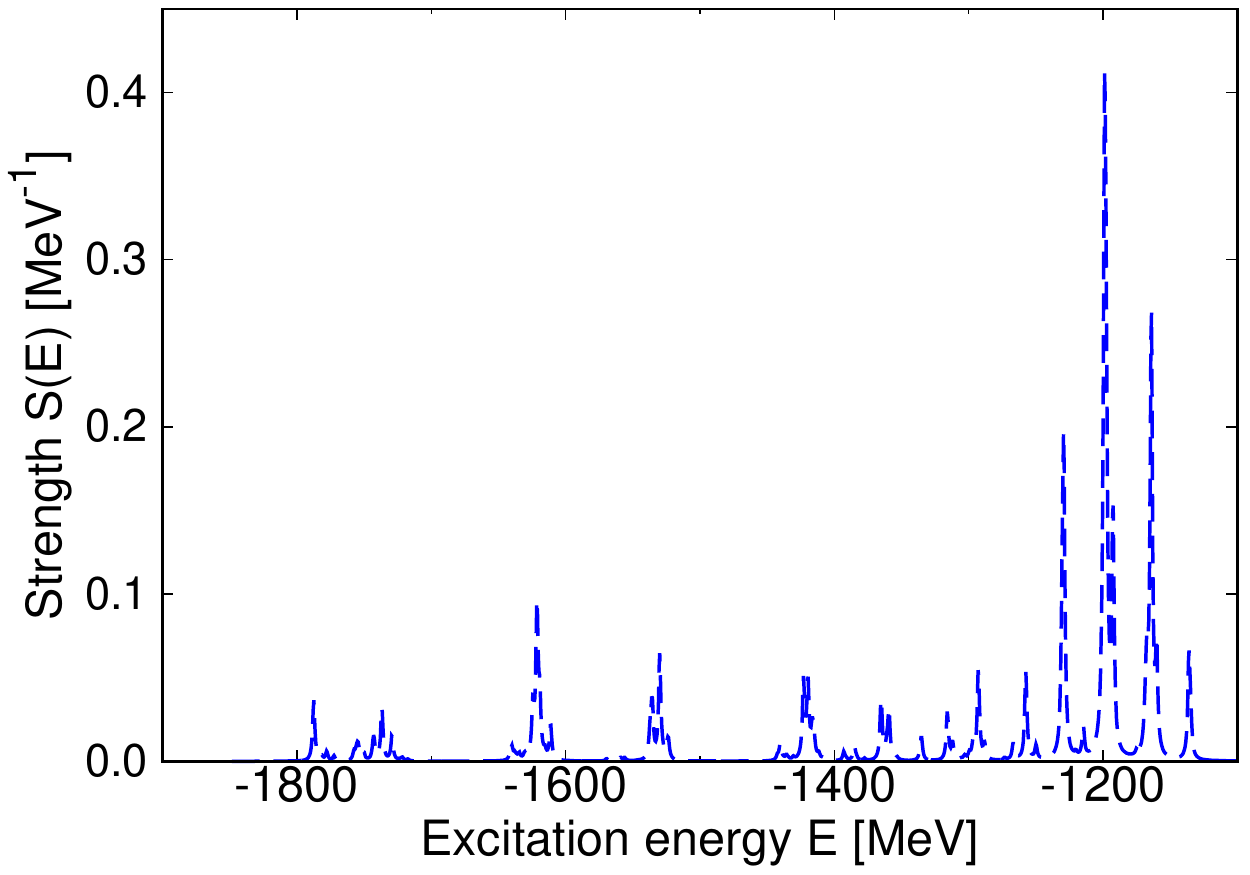}} \hfill%
\subcaptionbox{$^{78}$Ni GT$_-$ transition strength in the particle sector. \label{strength_78Ni_particle}}%
{\includegraphics[width=.49\textwidth] {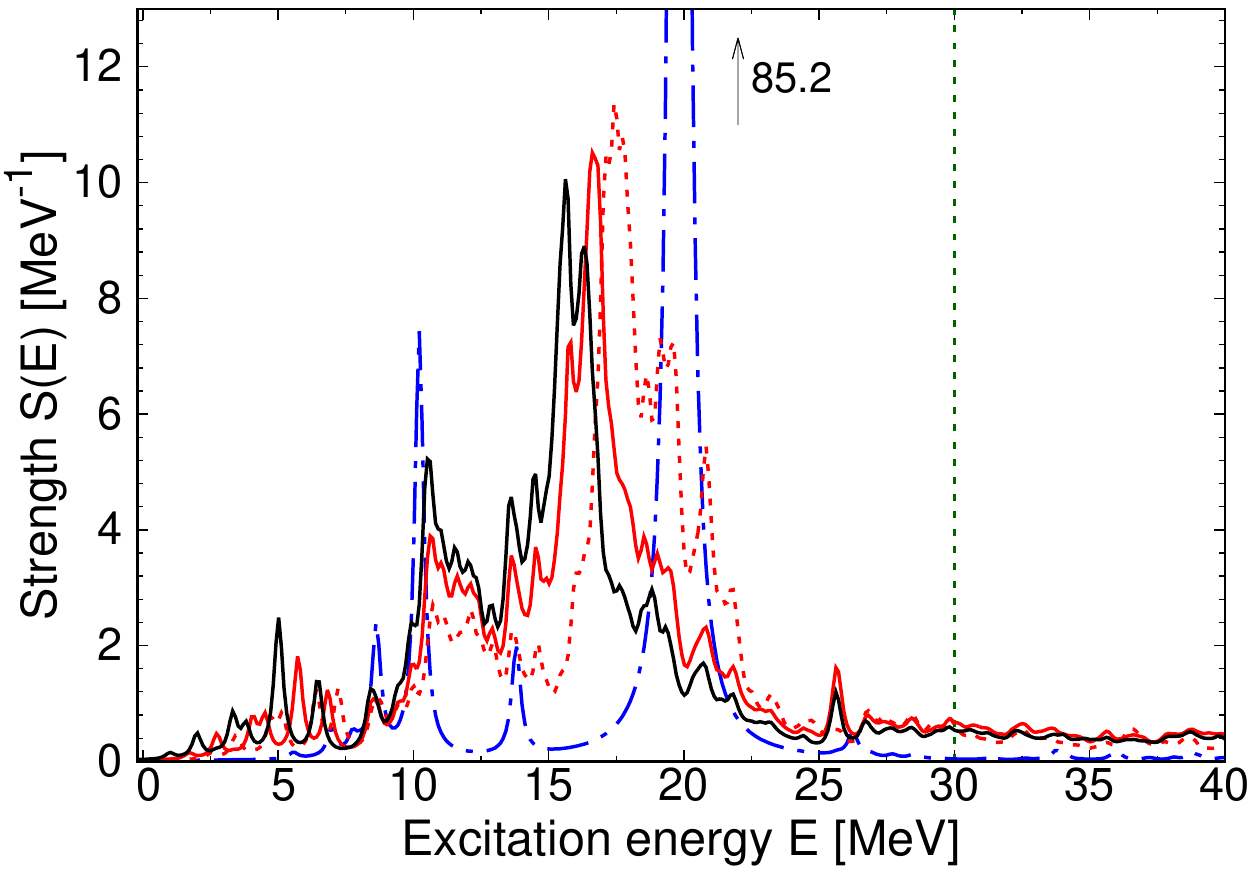}} \hfill%
\subcaptionbox{$^{78}$Ni cumulative sum of the B(GT$_-$) values. \label{strength_78Ni_cumu}}%
{\includegraphics[width=.6\textwidth] {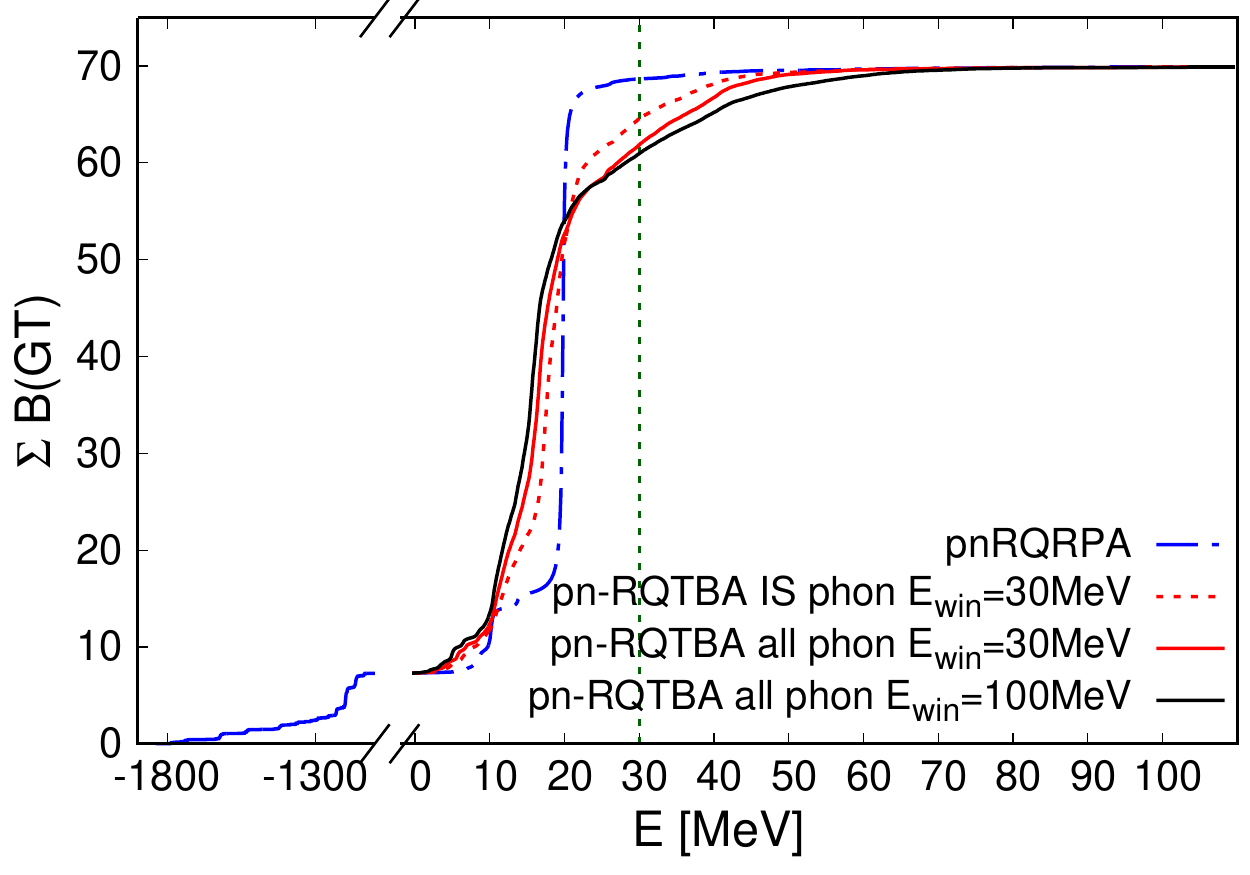}} \hfill %
\caption{GT$_-$ strength distribution in $^{78}$Ni and corresponding cumulative sum of B(GT$_-$). The results are shown at different theoretical levels: at the pn-RQRPA level (in blue), at the pn-RQTBA level with coupling to isoscalar (IS) phonons in a QVC energy window $E_{win}=30$ MeV in dashed red, at the pn-RQTBA level with coupling to IS and charge-exchange phonons in a QVC energy window $E_{win}=30$ MeV in plain red, and at the pn-RQTBA level with coupling to IS and charge-exchange phonons in a QVC energy window $E_{win}=100$ MeV in black.}
\label{fig:78Ni_2} 
\end{figure}

\begin{figure}[h!]
\centering
{\includegraphics[width=.5\textwidth] {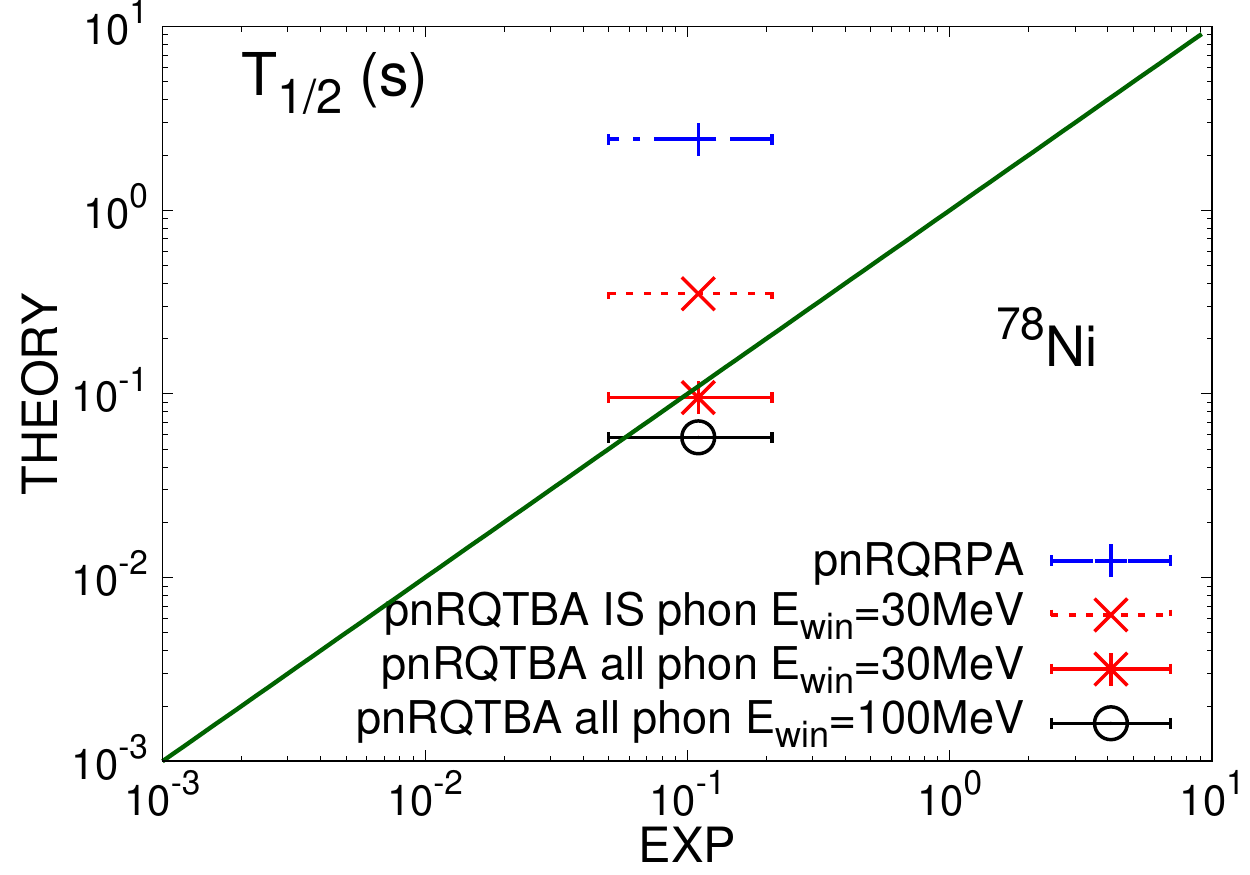}} 
\caption{Half-life of $^{78}$Ni in seconds. The experimental value is taken form Ref. \cite{Hosmer}. The error bars are experimental ones.}
\label{f:hlife}
\end{figure}

\begin{figure}
\centering
\subcaptionbox{GT$_-$ strength in $^{48}$Ca in the particle sector. \label{strength_48Ca_particle}}%
{\includegraphics[width=.49\textwidth] {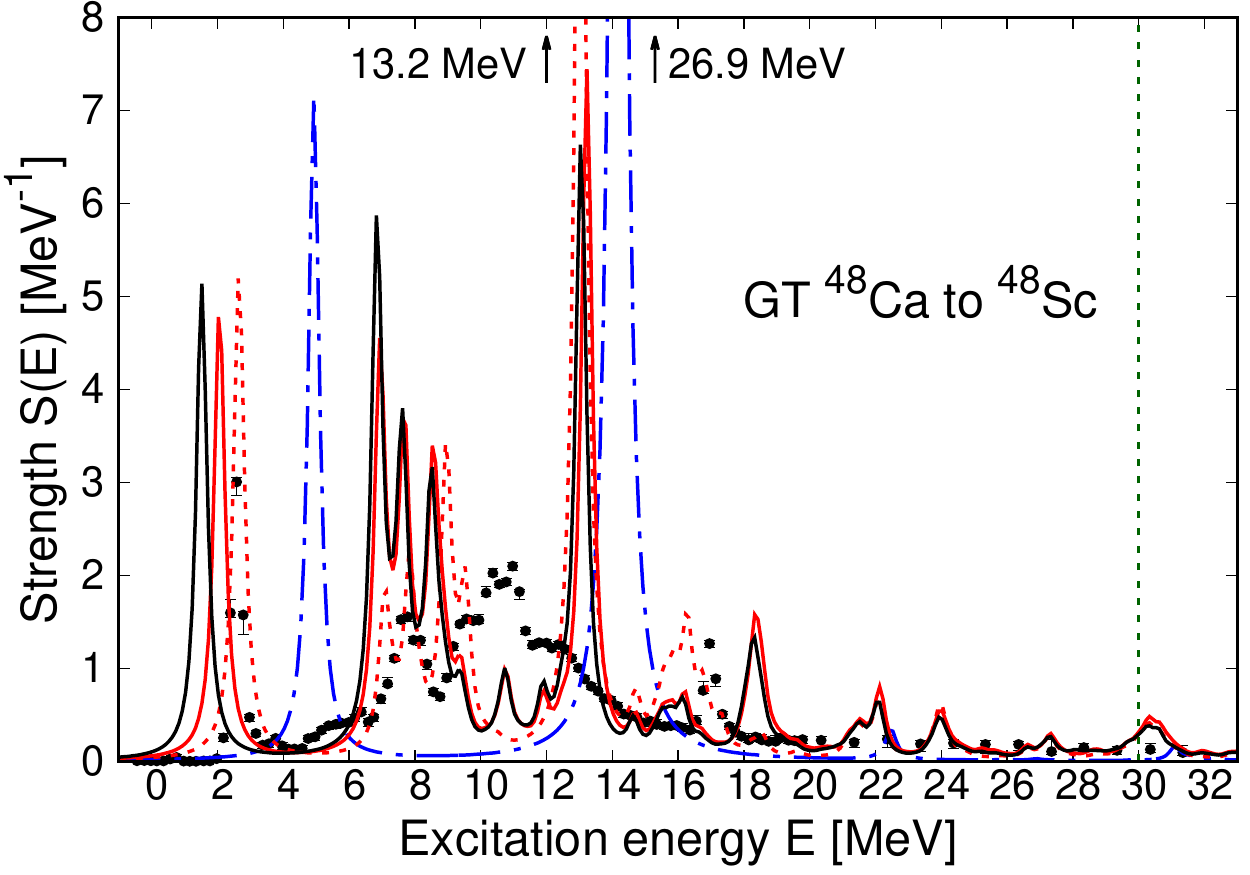}}  \hfill%
\subcaptionbox{$^{48}$Ca cumulative sum of the B(GT$_-$). \label{strength_48Ca_cumu}}%
{\includegraphics[width=.49\textwidth] {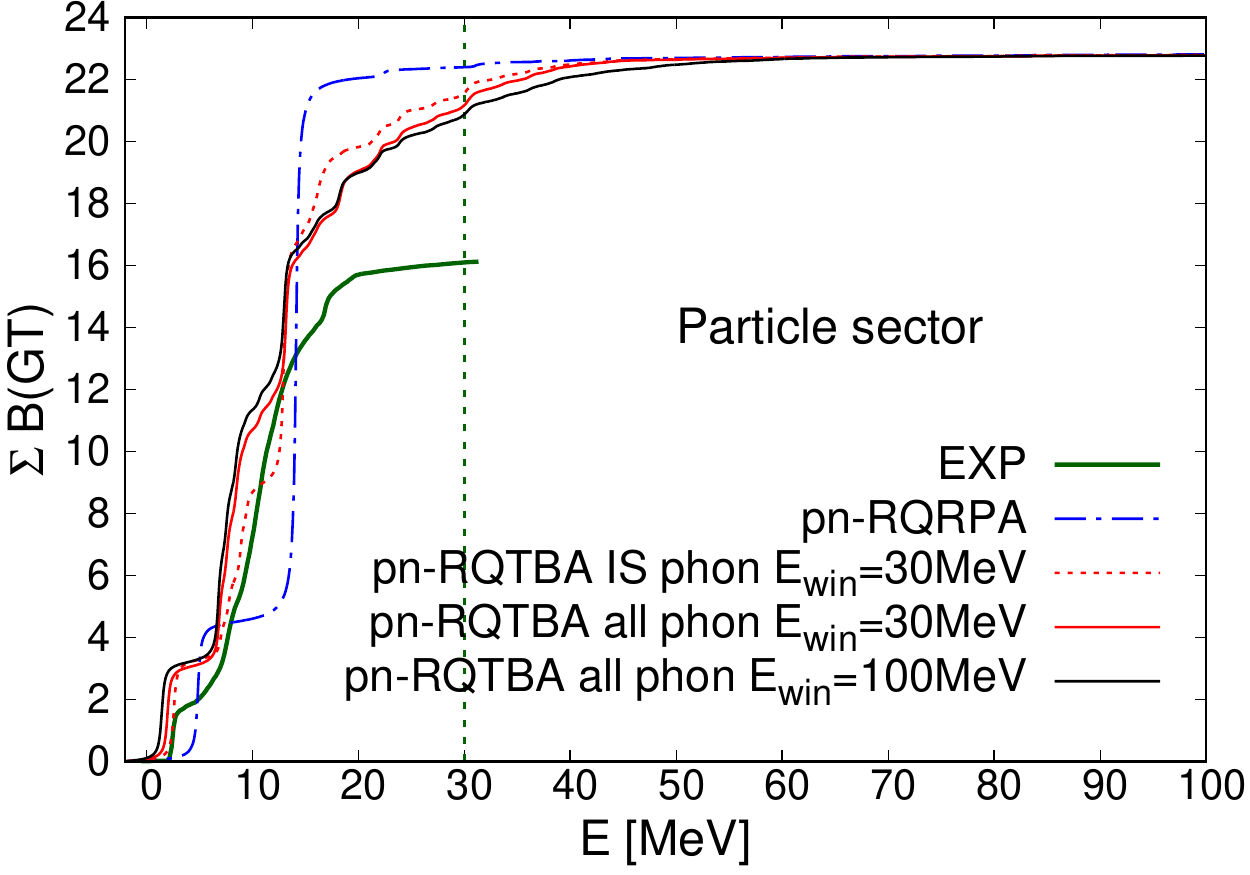}} \hfill %
\caption{GT$_-$ strength in $^{48}$Ca in the particle sector and cumulative sum of the B(GT$_-$) values.} 
\label{fig:48Ca_2} 
\end{figure}

Gamow-Teller (GT) modes describe the nuclear response to the following spin-isospin flip operator
\begin{equation}
\hat F^{GT}_{\pm} = \sum_{i=1}^A \boldsymbol{\Sigma}^{(i)} \tau_\pm^{(i)} \;, \hspace{0.2cm}  \mbox{with} \hspace{0.2cm}  
\boldsymbol{\Sigma}^{(i)} = 
\begin{pmatrix}
\boldsymbol{\sigma}^{(i)} & 0 \\
0 & \boldsymbol{\sigma}^{(i)} 
\end{pmatrix} \; .
\end{equation}
For such unnatural parity modes the pion gives almost the full contribution of the static meson exchange interaction (\ref{eq:int}). As it is absent in the RHB ground-state, the pion is considered with the free-space coupling constant $\frac{f_\pi^2}{4\pi} =  0.08$, which makes GT a great study-case where no double-counting effect is expected when going beyond the pn-RQRPA description.
In this work we calculate the response of stable and neutron-rich nuclei to the $\hat F^{GT}_-$ operator, using the numerical scheme described in Ref. \cite{Robin}.
Starting from the RHB approximation with the NL3 meson parametrization \cite{NL3} and a monopole-monopole pairing force \cite{Litvinova2008}, the set
of phonons that are coupled to the quasiparticles is calculated in the RQRPA.  
As a first step, we consider the coupling to non-isospin-flip phonons with $J^\pi=2^+,3^-,4^+,5^-,6^+$ in an energy window of $30$ MeV. The GT response is then calculated by solving Eq. (\ref{eq:BSE_pn}), and the strength distribution is obtained from Eq. (\ref{eq:strength}).
\\
In Ref. \cite{Robin} we calculated the GT$_-$ response of neutron-rich Nickel isotopes within this approach. In Figs. \ref{strength_68Ni_1} and \ref{strength_78Ni_1} we show the strength distributions in $^{68}$Ni and $^{78}$Ni calculated with a smearing parameter $\Delta = 200$ keV. For comparison we show the results obtained at the pn-RQRPA level in blue, while the full calculations including QVC (pn-RQTBA) are displayed with red lines. In order to obtain the excitation spectrum with respect to the daughter ground-state, all distributions have been shifted by the binding energy difference $\Delta B = B(parent) - B(daughter)$ calculated in the RHB approximation.
Clearly, the QVC induces fragmentation and spreading of the pn-RQRPA transition strength over a larger energy range including the low-energy region. 
We saw that such effects resulted in a considerable improvement of the $\beta$-decay half-lives as compared to the pn-RQRPA description \cite{Robin}. 
We also show in Fig. \ref{strength_68Ni_1} the low-lying strength distribution in $^{68}$Ni calculated with $\Delta=20$ keV in order to disentangle the low-lying states. Experimentally it is known that the ground-state of $^{68}$Cu has angular momentum and parity $1^+$ \cite{nndc}. The shift induced by QVC clearly improves the position of the $1^+_1$ state, as it is found to lie at $330$ keV versus $3.88$ MeV in pn-RQRPA. 
We remind that we do not introduce any static proton-neutron pairing. In pn-(R)QRPA studies such pairing is usually introduced through an attractive residual isoscalar pn interaction in the particle-particle channel which is fitted to \textit{e.g.} $\beta$-decay half-lives \cite{Engel,Mustonen,Bai,Niksic}. When the QVC interaction of Fig. \ref{fig:Phi_IS} is taken into account, pn pairing appears naturally due to the presence of particle-like pairing which generates the particle-particle component of the dynamical pn interaction $\Phi(\omega)$. This can explain why the QVC induces a similar effect on the low-energy strength as the phenomenological static pn pairing. \\ 
$\;$\\
\noindent We show in Figs. \ref{strength_90Zr_1} and \ref{strength_48Ca_1} the GT$_-$ strength distributions in $^{90}$Zr and $^{48}$Ca which have been measured experimentally.
The strength in $^{90}$Zr has been calculated using a smearing $\Delta = 1$ MeV, to simulate the experimental resolution of Ref. \cite{Wakasa}.
When QVC is included, a nice agreement with the data is obtained up to an excitation energy of $\sim25$ MeV. At higher energy the experimental strength also contains the isovector spin-monopole response \cite{Wakasa}.
The GT$_-$ strength in $^{48}$Ca has been measured with a higher energy resolution corresponding to $\Delta = 210$ keV in Ref. \cite{Yako}. At such high resolution, the pn-RQRPA is not able to reproduce the details of the experimental spectrum. The position of the first low-lying state and giant resonance region are also overestimated by $\sim 3-4$ MeV. 
The fragmentation induced by QVC leads to the appearance of finer details in the transition strength, which is also shifted down in accordance with the data. The peak around $\sim 13$ MeV however still lacks fragmentation.
\\ \\
\indent Up to now the phonon spectrum coupled to the quasiparticles was limited to non-isospin-flip phonons with natural parity. 
Ref. \cite{Litvinova2016} however showed that the presence of low-lying isospin-flip modes can have an impact on the single-nucleon shell structure. Recent developments now make it possible to include the coupling to charge-exchange phonons in the description of the response in doubly magic nuclei. Such vibrations introduce new diagrams in the QVC interaction that are displayed in Fig. \ref{fig:Phi_IS+IV}. We note that due to charge conservation they only generate additional self-energy insertions, and do not appear in the form of a phonon-exchange. Moreover, we see that these new self-energy terms involve proton-neutron particle-particle elements of the particle-vibration coupling vertex, and thus can also be interpreted as a (virtual) energy-dependent pn pairing interaction in doubly-magic nuclei.
$\;$ \\
\noindent Fig. \ref{strength_78Ni_particle} displays in plain red lines the GT$_-$ strength in $^{78}$Ni calculated when adding the coupling to charge-exchange vibrations with $J^\pi=0^\pm,1^\pm,2^\pm,3^\pm,4^\pm,5^\pm,6^\pm$. 
Although small, the effect of such phonons is noticeable as they slightly modify the giant resonance region, introduce a new state around 26 MeV and shift further down the low-energy part of the strength. The resulting half-life calculated with the effective weak axial vector coupling constant $g_A =1$ is shown in Fig. \ref{f:hlife} and is now in great accordance with the experimental value \cite{Hosmer}. 
In Fig. \ref{strength_78Ni_cumu}, we show the corresponding cumulative integrated strength, or cumulative sum of B(GT$_-$) values. In the particle sector, above the giant resonance region around 30 MeV we note that the pn-RQRPA strength has almost reached its saturated value ($97\%$). Introducing the coupling to isoscalar and isovector phonons (plain red curve) leads to a "quenching" of $\sim 16\%$ of the RQRPA strength at this energy due to fragmentation and redistribution of the strength.
We also note from Fig. \ref{strength_78Ni_Dirac} the presence of states at very large negative energy, corresponding to transitions to the Dirac sea of negative-energy states. As already mentioned in \cite{Paar}, such transitions are the result of the "empty Dirac-sea approximation" which is made at the relativistic mean-field level to avoid divergences. In the present case of $^{78}$Ni these transitions to the antiparticle sector take away about $10\%$ of the total GT$_-$ strength.
\\
In order to reproduce details of the excitation spectra, and to tackle the quenching problem of the GT strength, it is important to include complex nucleonic configurations in a model space which is as large as possible.
In the doubly magic nucleus $^{78}$Ni, it is possible to increase the QVC energy window, in which 1p-1h$\otimes$phonon excitations are included, up to $E_{win} = 100$ MeV that is the same energy cut-off which is used to select the 1p-1h pairs in pn-RQRPA. The final results are shown with black lines. The main effect of increasing the value of $E_{win}$ is an enhanced "quenching" at high excitation energy above 30 MeV, (Fig. \ref{strength_78Ni_cumu}), and a slight redistribution of the low-lying strength which leads to an additional decrease of the half-life (Fig. \ref{f:hlife}), that remains however within the experimental error bars. 
$\;$ \\
\noindent Finally we show in Fig. \ref{strength_48Ca_particle} and \ref{strength_48Ca_cumu} the GT$_-$ strength distribution in $^{48}$Ca in the particle sector and the corresponding cumulative sum of the B(GT$_-$) values.
The coupling to charge-exchange phonons brings further fragmentation of the peak at $\sim$ 13 MeV which is counterbalanced by an increase of the strength around $\sim$ 7 MeV. We now see a clear state at $\sim 18.3$ MeV which is $\sim 1.3$ MeV too high than the one seen experimentally. The first state around 2 MeV is shifted down when introducing charge-exchange phonons and increasing the QVC window $E_{win}$ to a value of 100 MeV.
Around 30 MeV, which is the excitation energy reached experimentally, we note again from Fig. \ref{strength_48Ca_cumu} that the pn-RQRPA distribution has almost saturated as it reaches $\sim 98\%$ of the total GT$_-$ strength in the particle sector. The strength with QVC (in black) introduces some quenching since it exhausts only $\sim 91 \%$ of that value. Such redistribution of the transition strength is however not sufficient to reach the experimental value of only $71\%$.

\section{Summary and outlook}
In this work we applied the pn-RQTBA approach to the description of Gamow-Teller (GT) transitions in a few mid-mass nuclei. We found that the coupling between quasiparticles and collectives vibrations, which introduces complex 1qp-1qh$\otimes$phonons configurations, systematically induces fragmentation and spreading of the pn-RQRPA transition strength, and leads to a more detailed description of the GT states, in better accordance with the data.
We included for the first time the coupling to charge-exchange vibrations in doubly magic nuclei and found that such phonons can have a non-negligible impact on the $\beta$-decay half-lives and on the quenching of the GT strength.
When the experimental resolution is high, it is however clear that the RQVC framework implemented in the present Time-Blocking Approximation is often not sufficient to reproduce the very fine details of the strength.
In order to further improve the description of spin-isospin modes we plan for the future to include higher-order configurations in the response \cite{Litvinova2015} and in the ground state.
\section{Acknowledgments}
We thank T. Marketin for providing part of the code for pn-RQRPA matrix elements. This work was supported by US-NSF Grants PHY-1404343 and PHY-1204486.

\end{document}